\documentclass[twocolumn,showpacs,preprintnumbers,amsmath,amssymb,superscriptaddress,prl]{revtex4}

\usepackage{graphicx}
\usepackage{graphics}
\usepackage{dcolumn}
\usepackage{bm}

\begin{document}

\title{Multiple charge density wave transitions in Gd$_2$Te$_5$}

\author{K. Y. Shin}
\affiliation{
Geballe Laboratory for Advanced Materials and Department of Applied Physics, Stanford University, Stanford, CA 94305 (USA)
}

\author{N. Ru}
\affiliation{
Geballe Laboratory for Advanced Materials and Department of Applied Physics, Stanford University, Stanford, CA 94305 (USA)
}

\author{C. L. Condron}
\affiliation{
Stanford Synchrotron Radiation Laboratory, Stanford Linear
Accelerator Center, 2575 Sand Hill Road, Menlo Park, California 94025,
USA
}

\author{Y. Q. Wu}
\affiliation{
Ames Laboratory and Department of Materials Science and Engineering, Iowa State University,Ames, IA 50011 (USA)
}

\author{M. J. Kramer}
\affiliation{
Ames Laboratory and Department of Materials Science and Engineering, Iowa State University,Ames, IA 50011 (USA)
}

\author{M. F. Toney}
\affiliation{
Stanford Synchrotron Radiation Laboratory, Stanford Linear
Accelerator Center, 2575 Sand Hill Road, Menlo Park, California 94025,
USA
}

\author{I. R. Fisher}
\affiliation{
Geballe Laboratory for Advanced Materials and Department of Applied Physics, Stanford University, Stanford, CA 94305 (USA)
}

\date{\today}

\begin{abstract}
Diffraction measurements performed via transmission electron microscopy and high resolution X-ray scattering reveal two distinct charge density wave transitions in Gd$_2$Te$_5$ at $T_{c1}$ = 410(3) and $T_{c2}$ = 532(3) K, associated with the \textit{on}-axis incommensurate lattice modulation and \textit{off}-axis commensurate lattice modulation respectively. Analysis of the temperature dependence of the order parameters indicates a non-vanishing coupling between these two distinct CDW states.
\end{abstract}


\maketitle

Charge density waves (CDWs) are a familiar concept in condensed matter physics\cite{Gruner}. 
The basic premise is that a large electronic susceptibility at finite wavevector $q$, such as can be generated by Fermi surface nesting in low dimensional materials, can lead to a coupled electronic/lattice instability if the electron-phonon coupling is strong enough. 
In recent years there has been renewed interest in this broken symmetry state, motivated in part by the desire to better understand charge-ordered states in strongly correlated systems\cite{Hong_2006}. 
The quasi two-dimensional rare earth ($R$) tellurides $R$Te$_2$ and $R$Te$_3$, based on single and double square Te planes respectively, have attracted particular interest given the combination of their simple electronic structure\cite{Kikuchi_1998,Laverock_2004,Shim_2004}, simple lattice modulation\cite{Fang_2007,Malliakas_2005,Kim_2006}, easy ``tunability'' (both in terms of band filling\cite{DiMasi_1995,DiMasi_1996} and also chemical pressure\cite{Ru_2008}), and finally the large magnitude of their CDW gap\cite{Brouet_2008}.
This renders the materials appropriate for several powerful and complementary probes of the electronic and physical structure\cite{Brouet_2004, Gweon_1998, Garcia_2007, Ru_2008a}. 
Several studies have established these as prototypical nesting-driven CDW systems, with CDW wave vectors determined by well-defined peaks in the Lindhard susceptibility\cite{Shin_2005,Hong_2006,Laverock_2004,Mazin_2008}.

Recently, the closely related family of compounds $R_2$Te$_5$ have also been shown to host CDW modulations \cite{Shin_2008, Malliakas_2008}.
Based on alternating single and double Te planes, interleaved by corrugated $R$Te layers\cite{DiMasi_1994}, this new family of compounds is essentially a hybrid of the simpler $R$Te$_2$ and $R$Te$_3$ compounds. 
This material raises the very interesting question of how CDW formation on the different Te planes coexists, or even competes. 
Our initial analysis indicated that CDWs originated independently on the single and double Te planes, driven by separate contributions to the susceptibility\cite{Shin_2008}. 
In this Rapid Communication, we establish via Transmission Electron Microscopy (TEM) and high resolution x-ray diffraction that the two sets of CDW wave vectors observed in Gd$_2$Te$_5$, one of which is incommensurate ($q_0$$\sim$0.69$c^*$), and the other of which is fully commensurate ($q_1$ =5/12$a^*$+1/12$c^*$, $q_2$ = 1/12$a^*$+5/12$c^*$), do indeed undergo separate CDW transitions, but that the two CDW condensates are not completely independent.

Single crystals of Gd$_2$Te$_5$ were grown by slowly cooling a binary melt, as described previously\cite{Shin_2008}. TEM diffraction images were taken at various temperatures along the (010) zone axis, i.e. perpendicular to the Te planes.
The samples were ion-milled using a liquid nitrogen cooling stage, and selected area diffration patterns (SADPs) were taken using a Philips CM 30 operating at 300keV with a Gatan double tilt heating stage for temperatures up to 540K, utilizing double copper washers to improve thermal contact. 
The equipment was optimized at the nominal camera length 900 mm and images were taken at varied exposure stops to obtain enough sensitivity for the weak superlattice reflections.
Real space images showed very few macroscopic defects, which were easily avoided.

\begin{figure}[tb]
\includegraphics[width=2.80in]{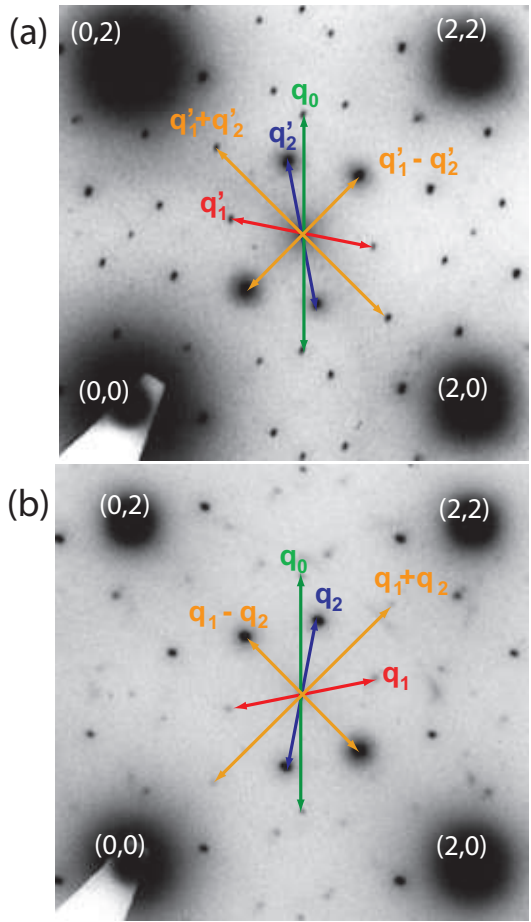}
\caption{(Color online) (a)Selected area TEM diffraction pattern in $(h0l)$ plane at room temperature (a) before and (b) after heating to 536 K. CDW wave vectors are labeled as described in the main text.}
\label{fig:TEMRoomT}
\end{figure}

SADPs taken at room temperature both before and after heating to 536 K are shown in Figure\ \ref{fig:TEMRoomT}, where the CDW wave vectors $q_0$, $q_1$ and $q_2$ are labeled around $(h0l)$=$(101)$. 
Figure\ \ref{fig:TEMRoomT}(a) shows that the initial diffraction pattern before heating is actually a mirror image of the satellite peaks previously observed in Ref.\onlinecite{Shin_2008} and we label the diffraction peaks accordingly $q'_1$= -5/12$a^*$+1/12$c^*$ and $q'_2$= -1/12$a^*$+5/12$c^*$.
After heating and subsequent cooling to room temperature(Figure\ \ref{fig:TEMRoomT}(b)), the diffraction pattern shows weak circular streaks, caused by irreversible surface recrystallization at high temperatures.
The diffraction pattern has also suffered a mirror reflection about the $c^*$ axis, ascribed to a reversal of the CDW domain orientation.

\begin{figure}[tb]
\includegraphics[width=2.95in]{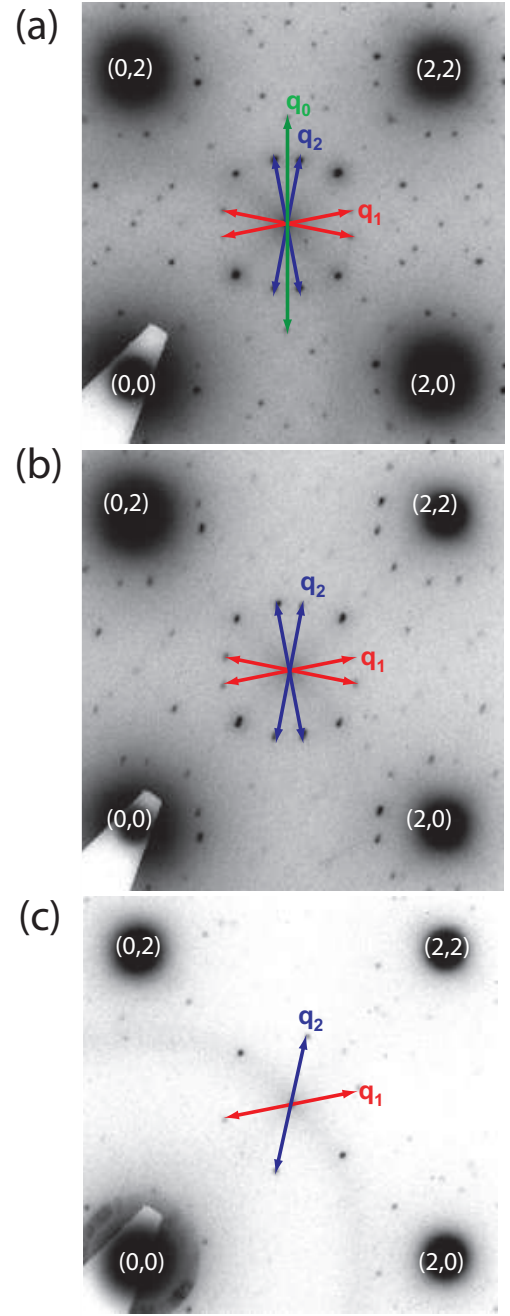}
\caption{(Color online) SADPs at (a) 343 K (b) 446 K and (c) 536 K. Wave vectors are labeled as described in the main text. The incommensurate CDW with $q_0$ $\sim$ 0.69 $c*$ has disappeared by 446 K, suggestive of a CDW transition below this temperature. 
}
\label{fig:TEMHighT}
\end{figure}

Representative SADPs taken at 343 K, 446 K and 536 K are shown in Figure\ \ref{fig:TEMHighT} in a sequential order of temperature changes. 
Both domains of the commensurate CDW (i.e. $q_1$ and $q_2$ oriented to the right and to the left) were observed for temperatures above 313K (see for example Figure\ \ref{fig:TEMHighT} (a)). 
Compared to the commensurate \textit{off}-axis CDW peaks, the incommensurate CDW along the $c^*$ axis did not show a similar effect, due to the inequivalence of the $a$ and $c$ axes.

The diffraction intensities for $q_0$ were tremendously reduced by 343 K (green arrows in Figure\ \ref{fig:TEMHighT} (a)) and had disappeared by 446 K (Figure\ \ref{fig:TEMHighT} (b)).  
In contrast, the intensities for the \textit{off}-axis CDW remained strong until much higher temperatures, eventually almost vanishing at the highest temperatures (536 K, Figure\ \ref{fig:TEMHighT} (c)).

\begin{figure}[tbp]
\includegraphics[width=3.3in]{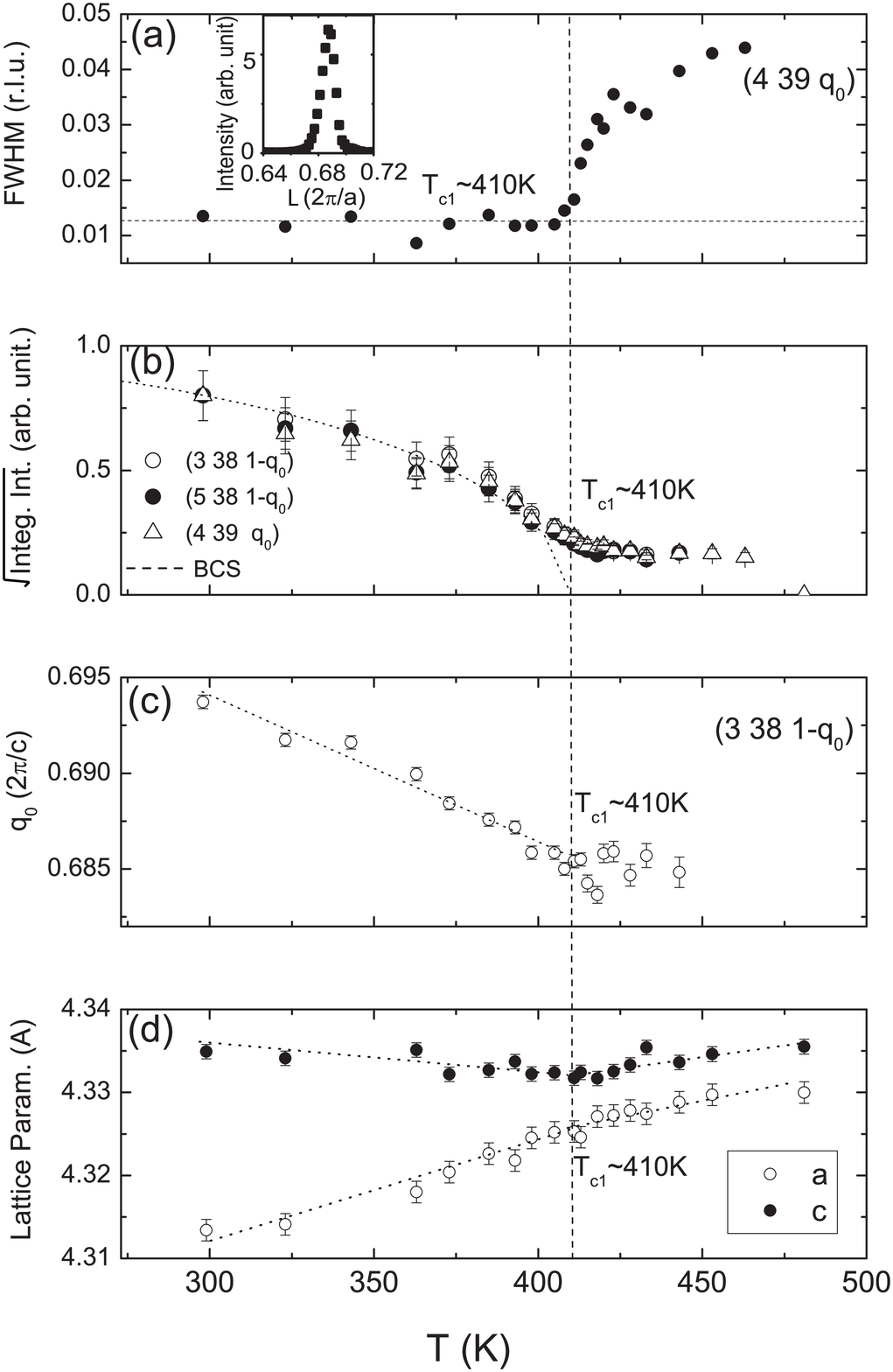}
\caption{ 
Temperature dependence of the \textit{on}-axis incommensurate CDW state. 
(a) FWHM of the CDW peak at (4 39 $q_0$) in the $ac$ plane.
Inset shows L scan at 300 K in the inset.
The sudden increase in FWHM at $T_{c1}$ = 410(3) K indicates a CDW phase transition. 
(b) Square root of the integrated intensity for peaks at (3 38 1-$q_0$), (5 38 1-$q_0$) and (4 39 $q_0$), normalized to the BCS curve at 300 K.
(c) CDW wave vector $q_0$ as a function of temperature measured from the CDW peak at (3 38 1-$q_0$). 
(d) In-plane lattice parameters $a$ and $c$ as a function of temperature.
The dashed vertical line for all panels indicates the nominal transition temperature $T_{c1}$ = 410 K.
}
\label{fig:incommenQT}
\end{figure}

The CDW transitions were investigated in greater detail via high resolution x-ray diffraction. 
Samples were glued on the surface of the metallic sample stage using silver epoxy and the temperature was actively controlled by an Anton-Paar furnace up to 550 K.
The samples were kept in a helium gas flow during the entire experiment in order to minimize oxidation.
Measurements were carried out in a reflection geometry for photon energies 9.35keV and 12.70keV at the Stanford Synchrotron Radiation Laboratory (SSRL) on Beam Lines 11-3 and 7-2.
A Ge(111) crystal analyzer or either 1 or 2 milliradian slits was selected depending on the measurement.
Bragg peaks and the CDW satellite peaks over a wide range of $(hkl)$ were carefully inspected in order to select regions of minimal structural defects.
Samples had to be realigned at each temperature due to the thermal expansion of the sample stage and the Bragg peaks near to the satellite peaks were centered at the maximum intensities.
CDW peaks are sharply peaked(inset Figure\ \ref{fig:incommenQT} (a)) 
and we obtain a lower bound for the CDW correlation length of $\sim 0.5\mu m$ in the $ac$ plane and $ \sim 0.05\mu m$ along the $b$-axis for both \textit{on}- and \textit{off}-axis CDWs.

The temperature dependence of the incommensurate CDW peaks for $(hkl)$=(3 38 1-$q_0$), (5 38 1-$q_0$) and (4 39 $q_0$) is shown in Figure\ \ref{fig:incommenQT}. 
A rapid increase in FWHM above the resolution limit (shown in Figure\ \ref{fig:incommenQT} (a) for the specific peak (4 39 $q_0$) in the $ac$ plane) signals a CDW transition at $T_{c1}$ = 410(3) K, consistent with TEM data described above.

The square root of the integrated intensity, proportional to the order parameter for weakly coupled systems, is shown in Figure\ \ref{fig:incommenQT} (b) for all three peaks, together with the mean field BCS curve drawn for $T_{c1}$ = 410 K. 
Data have been normalized to the BCS curve at 300 K.  
Residual scattering intensity above $T_{c1}$, ascribed to fluctuations due to the large FWHM we observe, was observed up to 463 K, above which the satellite peaks became too broad to be distinguished from background.

The absolute value of $q_0$ changes with temperature, increasing by approximately 1.5\% from $T_{c1}$ to room temperature (Figure\ \ref{fig:incommenQT}(c)), and indicating a fully incommensurate CDW. 
Above $T_{c1}$, $q_0$ does not appear to vary as strongly with temperature, but the accuracy of the measurement was limited by the significant broadening of the CDW peak.

The CDW transition is also apparent in the lattice parameters (Figure\ \ref{fig:incommenQT}(d)). 
Above $T_{c1}$, there is only a small difference (approximately 0.15\%) in the in-plane lattice parameters $a$ and $c$. 
On cooling below $T_{c1}$, there is a marked change in the thermal expansion coefficients, with the $a$-axis lattice parameter decreasing more rapidly with reducing temperature, while the $c$-axis lattice parameter actually increases with reducing temperature, at least in the range from 410 down to 300 K. 
By room temperature, the $c$-axis lattice parameter is fully 0.60\% larger than the $a$-axis. 
A similar ``stretching'' of the c-axis upon CDW formation was also observed for the unidirectional incommensurate CDW in TbTe$_3$\cite{Ru_2008}.

\begin{figure}[tbp]
\includegraphics[width=3.3in]{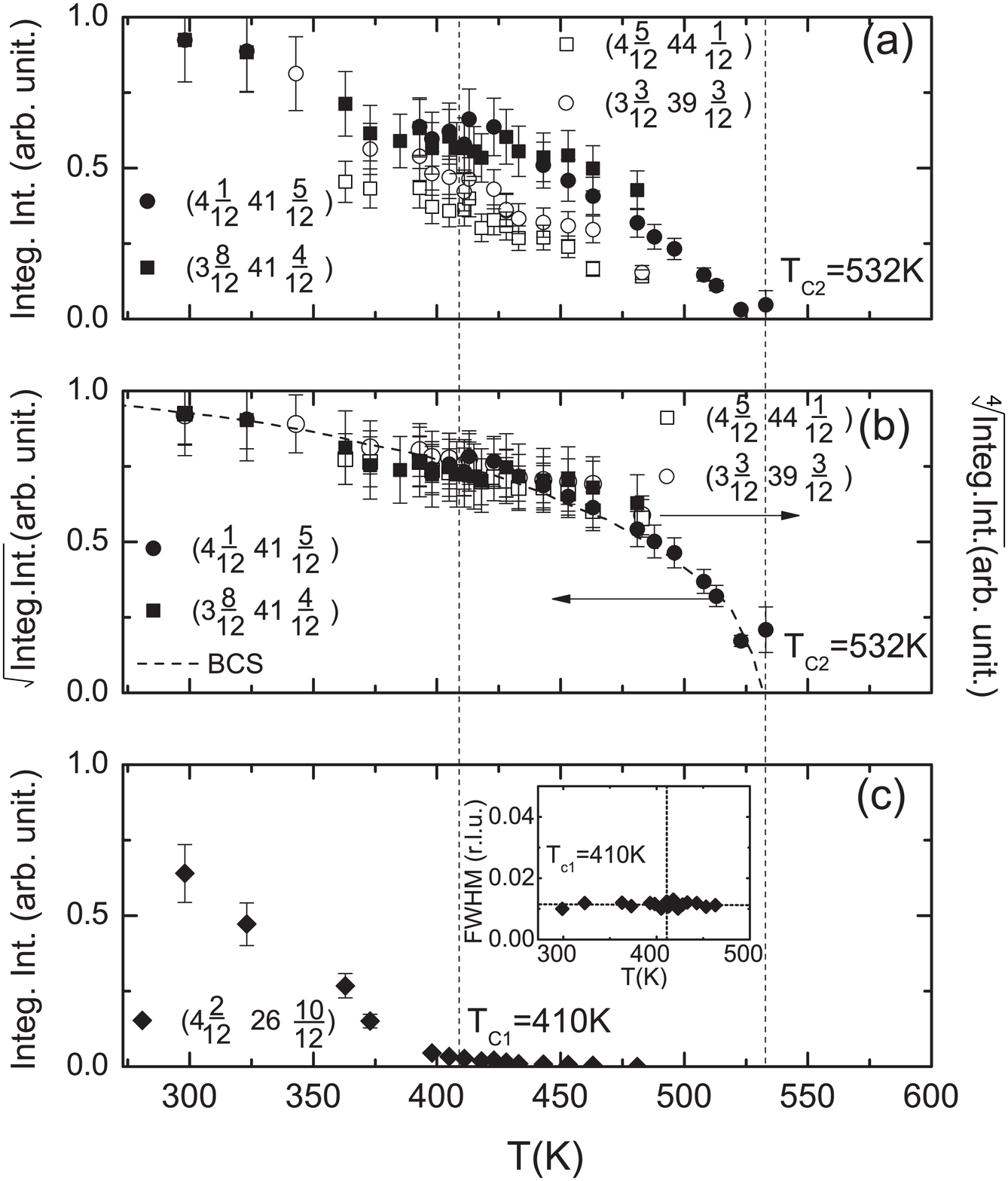}
\caption{Temperature dependence of the \textit{off}-axis commensurate CDW state. 
(a) Integrated intensities of CDW peaks at (4$\frac{1}{12}$ 41 $\frac{5}{12}$),(3$\frac{8}{12}$ 41 $\frac{4}{12}$), (4$\frac{5}{12}$ 44 $\frac{1}{12}$) and (3$\frac{3}{12}$ 39 $\frac{3}{12}$).
(b) Square roots of the integrated intensities of (4$\frac{1}{12}$ 41 $\frac{5}{12}$) and (3$\frac{8}{12}$ 41 $\frac{4}{12}$) (left axis) and 4th roots of the integrated intensities of (4$\frac{5}{12}$ 44 $\frac{1}{12}$) and (3$\frac{3}{12}$ 39 $\frac{3}{12}$) (right axis).
(c) Integrated intensity of the CDW peak at (4$\frac{2}{12}$ 26 $\frac{10}{12}$). 
Inset shows FWHM of the CDW peak at (4$\frac{2}{12}$ 26 $\frac{10}{12}$) in the $ac$ plane.
The intensity decreases significantly at $T$ $\sim$ $T_{c1}$, suggestive of interaction between the \textit{on}-and \textit{off}-axis CDWs.
The BCS order parameter is shown by a dashed line in panel (b), and vertical dashed lines indicate $T_{c1}$ = 410 K and $T_{c2}$ = 532 K. 
}
\label{fig:commenOrderP}
\end{figure}

The temperature dependence of the \textit{off}-axis commensurate CDW diffraction peaks was also measured. 
Data were collected for many peaks and representative measurements for five specific wavevectors are shown in Figure\ \ref{fig:commenOrderP}. 
The diffraction intensities for (4$\frac{1}{12}$ 41 $\frac{5}{12}$), (3$\frac{8}{12}$ 41 $\frac{4}{12}$), (4$\frac{5}{12}$ 44 $\frac{1}{12}$) and (3$\frac{3}{12}$ 39 $\frac{3}{12}$) decreased upon heating and almost disappeared at 533K (Figure\ \ref{fig:commenOrderP} (a)). 
Oxidation at these elevated temperatures, even in the flowing He atmosphere, meant that data had to be collected rapidly.  
Consequently, measurements of the FWHM were limited to temperatures below 533 K and, in contrast to the \textit{on}-axis CDW described above, it was difficult to systematically determine a sharp increase of the peak width associated with the ultimate CDW transition. 
However, comparison of the temperature dependence of the square roots of the integrated intensities of (4$\frac{1}{12}$ 41 $\frac{5}{12}$) and (3$\frac{8}{12}$ 41 $\frac{4}{12}$) and the 4th roots of the integrated intensities of (4$\frac{5}{12}$ 44 $\frac{1}{12}$) and (3$\frac{3}{12}$ 39 $\frac{3}{12}$) with the classical BCS order parameter (Figure\ \ref{fig:commenOrderP} (b)) allows an estimate of $T_{c2}$ = 532(3) K for these peaks, in reasonable agreement with the trend observed in TEM. 
This comparison implies that the satellite peaks at (4$\frac{1}{12}$ 41 $\frac{5}{12}$) and (3$\frac{8}{12}$ 41 $\frac{4}{12}$), corresponding to $q_2$ and $q_2$-$q_1$ respectively are from the fundamental, whereas peaks at (4$\frac{5}{12}$ 44 $\frac{1}{12}$), corresponding to $q_1$ and (3$\frac{3}{12}$ 39 $\frac{3}{12}$), corresponding to $q_2$-$2q_1$ are second harmonics, and hence that the fundamental wavevectors associated with the \textit{off}-axis CDW are actually $q_2$ = $\frac{1}{12}a^*$+$\frac{5}{12}c^*$ and $q_2$-$q_1$ = $-\frac{4}{12}a^*$+$\frac{4}{12}c^*$. 
This assignation might be consistent with the relative intensities of the superlattice peaks seen in Figure\ \ref{fig:TEMRoomT}, but it is not clear that this is the only interpretation for such a fully commensurate lattice modulation.

Inspection of the data in Figures\ \ref{fig:commenOrderP} (a) and (b) indicates that the intensity of these commensurate superlattice reflections might be slightly affected at the onset of the incommensurate \textit{on}-axis CDW at $T_{c1}$, but this effect is clearly at the limit of our resolution for these peaks. 
However, other commensurate CDW peaks exhibit a stronger effect. 
For example, the strongest suppression of the reflections for the commensurate superlattice at $T_{c1}$ occurred for the CDW peak at (4$\frac{2}{12}$ 26 $\frac{10}{12}$), where the intensity was suppressed almost to zero without increase in the FWHM up to much higher temperatures, indicating the absence of an actual phase transition associated with this specific wavevector (Figure\ \ref{fig:commenOrderP}(c) and the inset therein).
This behavior could not be accounted for by any order of harmonic generation, and suggests a non-trivial coupling of the commensurate and incommensurate CDW order parameters. 
It will require further experiments to elucidate the nature of this coupling, and the extent to which it affects the underlying electronic structure.

In summary, we have established that the \textit{on}-axis incommensurate and \textit{off}-axis commensurate lattice modulations in Gd$_2$Te$_5$ have different CDW transitions. 
Driven separately by different sections of the Fermi surface\cite{Shin_2008}, these two CDWs are not completely independent and
weak coupling between them results in deviation from BCS temperature dependence in some superlattice peaks associated with the commensurate CDW state.

\section{\label{sec:acknowlodegements}acknowledgements}

This work is supported by the DOE, Office of Basic Energy Sciences, under Contract No. DE-AC02-76SF00515. 
Efforts at the Ames Laboratory were supported by the DOE under Contract No. DE-AC02-07CH11358.
KYS was partly supported by the HS Lee Foundation in South Korea.
Portions of this research were carried out at the Stanford Synchrotron Radiation Laboratory(SSRL), a national user facility operated by Stanford University on behalf of the US Department of Energy, Office of Basic Energy Sciences.


\end{document}